# STATISTICAL RANDOM NUMBER GENERATOR ATTACK AGAINST THE KIRCHHOFF-LAW-JOHNSON-NOISE (KLJN) SECURE KEY EXCHANGE PROTOCOL[1]

CHRISTIANA CHAMON[*], SHAHRIAR FERDOUS, and LASZLO B. KISH

*Department of Electrical and Computer Engineering*
*Texas A&M University*
*400 Bizzell St*
*College Station, TX 77843*

This paper introduces and demonstrates four new statistical attacks against the Kirchhoff-Law-Johnson-Noise (KLJN) secure key exchange scheme. The attacks utilize compromised random number generators at Alice's/Bob's site(s). The case of partial correlations between Alice's/Bob's and Eve's probing noises is explored, that is, Eve's knowledge of Alice's and Bob's noises is limited but not zero. We explore the bilateral situation where Eve has partial knowledge of Alice's and Bob's random number generators. It is shown that in this situation Eve can crack the secure key bit by taking the highest cross-correlation between her probing noises and the measured voltage noise in the wire. She can also crack the secure key bit by taking the highest cross-correlation between her noise voltages and her evaluation of Alice's/Bob's noise voltages. We then explore the unilateral situation in which Eve has partial knowledge of only Alice's random number generator thus only those noises (of Alice and Eve) are correlated. In this situation Eve can still crack the secure key bit, but for sufficiently low error probability, she needs to use the whole bit exchange period for the attack. The security of the KLJN key exchange scheme, similarly to other protocols, necessitates that the random number generator outputs are truly random for Eve.

## 1. Introduction[1]

### 1.1. *On Secure Communications*

One way to establish the security of a communication is through encryption, that is, the conversion of plaintext into ciphertext via a cipher [1]. Fig. 1 provides the general scope of symmetric-key cryptography [1]. The key is a shared secret (string of random bits). Both communicating parties Alice and Bob use the same key and ciphers to encrypt and decrypt their plaintext [2, 3].

---

[*]Corresponding author

[1]Copied and adapted from [2, 3]

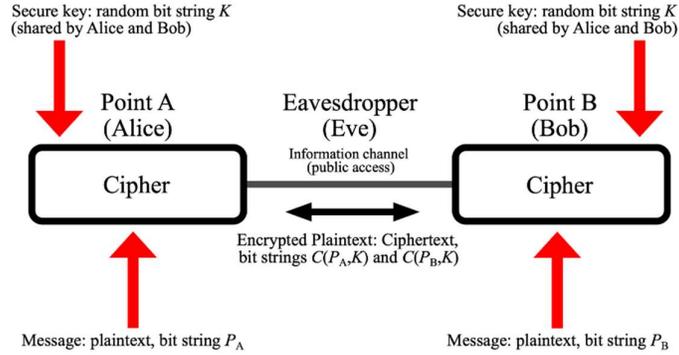

Fig. 1. Symmetric-key cryptography [1]. Alice and Bob securely exchange a key (a string of random bits) through the information channel. The ciphers encrypt plaintext into the ciphertext $C$. The secure key is $K$, the plaintext messages of Alice and Bob are denoted by $P_A$ and $P_B$, respectively, and the ciphertext is a function $C(P,K)$ [2, 3].

For a plaintext message $P$ and a secure key $K$, the encrypted message, or the ciphertext $C$, is a function of $P$ and $K$, that is,

$$C = C(P, K) \, . \tag{1}$$

In symmetric-key cryptography, for decryption, the inverse operation is used [2, 3]:

$$P = C^{-1}[C(P,K), K] \, . \tag{2}$$

Because the secure keys must be the same at the two sides (shared secret), another type of secure data exchange is needed before the encryption protocol can begin. It is the secure key exchange, which is the generation and distribution of the secure key over the communication channel. Usually, this is the most demanding process in secure communications because the communication channel is accessible by Eve thus the secure key exchange is itself a secure communication.

Eve records the whole communication during the key exchange, too. She knows every detail of the devices, protocols, and algorithms in the permanent communication system (as stated by Kerckhoffs's[2] principle [4]/Shannon's maxim), except for the secure key. In the ideal case of perfect security, the key is securely generated/shared, immediately used by a One Time Pad [5], and destroyed after the usage. In practical cases, usually there are deviations from these strict conditions, yet the general rule holds: A secure system cannot be more secure than its key [2, 3].

The key is assumed to be generated from truly random numbers. Any predictability of the key leads to compromised security [5]. In this paper, we demonstrate attacks on the unconditionally secure Kirchhoff-Law-Johnson-Noise (KLJN) secure key exchange based on compromised random number generators (RNGs), implying compromised noises of the communicating parties.

---

[2]Auguste Kerckhoffs, not to be confused with Gustav Kirchhoff

### 1.2. *On the KLJN scheme*

The KLJN system [2, 3, 6–61] is a statistical physical scheme based on the thermal noise of resistors. It is a classical (statistical) physical alternative of Quantum Key Distribution (QKD). As an illustration of the difficulties and depth of the issues of security, in papers [5, 62–98], important criticisms and attacks are presented about QKD, indicating some of the most important aspects of unconditionally secure quantum hardware and their theory [2, 3].

Fig. 2 illustrates the core of the KLJN scheme. The two communicating parties, Alice and Bob, are connected via a wire. They have identical pairs of resistors, $R_A$ and $R_B$. The statistically independent thermal noise voltages, $U_{H,A}(t)$, $U_{L,A}(t)$, and $U_{H,B}(t)$, $U_{L,B}(t)$, represent the noise voltages of the resistors $R_H$ and $R_L$ ($R_H > R_L$) of Alice and Bob, respectively. These are often generated from random number generators (RNGs) [2, 3] and must have a Gaussian amplitude distribution [21, 24].

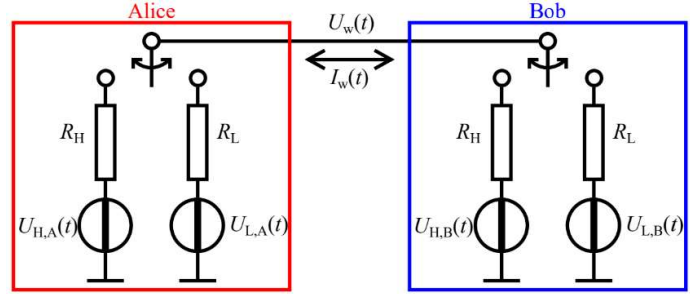

Fig. 2. The core of the KLJN scheme. The two communicating parties, Alice and Bob, are connected via a wire. The wire voltage and current are denoted as $U_w(t)$ and $I_w(t)$, respectively. The parties have identical pairs of resistors $R_H$ and $R_L$ ($R_H > R_L$) that are randomly selected and connected to the wire at the beginning of the bit exchange period. The statistically independent thermal noise voltages $U_{H,A}(t)$, $U_{L,A}(t)$, and $U_{H,B}(t)$, $U_{L,B}(t)$ represent the noise voltages of the resistors $R_H$ and $R_L$ of Alice and Bob, respectively [2, 3].

At the beginning of each bit exchange period (BEP), Alice and Bob randomly choose one of their resistors to connect to the wire. The wire voltage $U_w(t)$ and current $I_w(t)$ are as follows:

$$U_w(t) = I_w(t) R_B + U_B(t),\qquad(3)$$

$$I_w(t) = \frac{U_A(t) - U_B(t)}{R_A + R_B}\qquad(4)$$

where $U_A(t)$ and $U_B(t)$ denote the instantaneous noise voltage of the resistor chosen by Alice and Bob, respectively. Alice and Bob (as well as Eve) use the mean-square voltage on the wire to assess the situation. According to the Johnson formula,

$$U_w^2 = 4kT_{eff} R_p \Delta f_B,\qquad(5)$$

where $k$ is the Boltzmann constant ($1.38 \times 10^{-23}$ J/K), $T_{\text{eff}}$ is the publicly agreed effective temperature, $R_P$ is the parallel combination of Alice's and Bob's connected resistors, given by

$$R_P = \frac{R_A R_B}{R_A + R_B},\tag{6}$$

and $\Delta f_B$ is the noise bandwidth of the generators [2, 3].

Four possible resistance combinations can be formed by Alice and Bob: HH, LL, LH, and HL. Using the Johnson formula (Equation 5), these correspond to three mean-square voltage levels, as shown in Fig. 3.

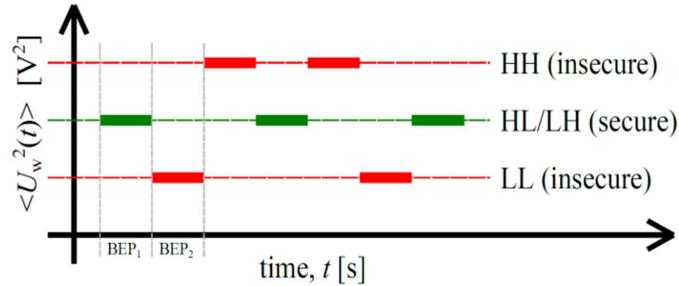

Fig. 3. The three mean-square voltage levels. The HH and LL cases represent insecure situations because they form distinct mean-square voltages. The HL and LH cases represent secure bit exchange because Eve cannot distinguish between the corresponding two resistance situations (HL and LH). On the other hand, Alice and Bob can determine the resistance at the other end because they know their own connected resistance value [2, 3].

The HH and LL cases represent insecure situations because they form distinct mean-square voltages. The HL and LH cases represent secure bit exchange because Eve cannot distinguish between the corresponding two resistance situations (HL and LH). On the other hand, Alice and Bob can determine the resistance at the other end because they know their own connected resistance values [2, 3].

Several attacks against the KLJN system have been proposed [2, 3, 14, 41–61], but no attack has been able to compromise its information-theoretic security. Each of these attacks is either invalid or it can be nullified by a corresponding defense scheme. In other words, by using the proper protocols, Eve's information entropy about the key approaches the bit length of the key, while Alice's and Bob's information entropy about the key approaches zero. The attacks presented in this paper are based on the assumption that the random number generators Alice and Bob use to generate their noises are compromised.

### 1.3. *On random number generator attacks*

There are two classes of practical random number generators: true (physical) and computational. The nature of computational RNGs is that they collect randomness from various low-entropy input streams and try to generate outputs that are practically indistinguishable from truly random streams [99–104]. The randomness of an RNG relies on the uncertainty of the random seed, or initialization vector, and a long sequence with uniform distribution.

The moment an adversary learns the seed, the outputs are known, and the RNG is compromised [2, 3].

Various RNG attacks exist against conditionally secure communications [94–99]. Unconditionally secure communications also require true random numbers for perfect security, and that is also true for the noises of Alice and Bob and for the randomness of their switch driving, which uses a different RNG from the voltage sources. The RNG outputs must imitate thermal noise, and in many practical applications, only a computational RNG is possible.

Chamon, et. al. published a paper on an RNG attack against the KLJN scheme [2, 3]. The attack was a deterministic attack using probing noises that exactly match the noises of Alice and Bob. In this paper, we demonstrate that even statistical knowledge of Alice's/Bob's noises can lead to significant information leak when Eve uses proper attack protocols.

The rest of this paper is organized as follows. Section 2 describes two new attack protocols, Section 3 demonstrates the results, and Section 4 concludes this paper.

## 2. Statistical Attack Protocol

Two situations are introduced where Eve can use compromised RNGs to crack the KLJN scheme: one where Eve has partial knowledge of both Alice's and Bob's generators (bilateral knowledge), and another where Eve has partial knowledge of only Alice's generator (unilateral knowledge).

### 2.1. *Bilateral knowledge*[3]

Eve has statistical knowledge of the amplitudes of the noise voltage generators $U_{H,A}(t)$, $U_{L,A}(t)$, $U_{H,B}(t)$, and $U_{L,B}(t)$ for each of Alice's and Bob's resistors, see Fig. 2. These noises are correlated with Alice's and Bob's corresponding noises. Eve then uses Equations 3 and 4 for the cross-correlation attacks shown below.

#### 2.1.1. *Cross-correlation attack utilizing Alice's/Bob's and Eve's channel voltages, currents and power*

For the four possible resistor combinations, HL, LH, LL and HH, Eve sets up a simulator utilizing her noises (that are correlated with Alice's Bob's corresponding noises), and she records the resulting wire voltages $U_{HH}(t)$, $U_{LL}(t)$, $U_{HL}(t)$, and $U_{LH}(t)$. Then she evaluates the cross-correlation between the *measured* wire voltage $U_w(t)$ of Alice and Bob, and the simulated ones. For example, in the HH simulated resistor situation, this is given by

$$CCC_{HH,U} = \frac{\langle U_w(t)U_{HH}(t)\rangle}{U_w U_{HH}},$$ (7)

where $CCC_{HH,U}$ is the cross-correlation coefficient between her simulated wire voltage $U_{HH}(t)$ in the HH case and the *measured* channel voltage $U_w(t)$.

---

[3]Copied and adapted from [3]

Finally, Eve guesses that the actual resistor situation is the one with the highest cross-correlation between the voltage on the wire and in Eve's probing system.

Note, this protocol of cross-correlating the simulated and real wire data can also be repeated for the channel current $I_w(t)$ and channel power $P_w(t)$.

For the channel current protocol, she measures $I_w(t)$ and evaluates the cross-correlation between the measured $I_w(t)$ and her four simulated wire currents $I_{HH}(t)$, $I_{LL}(t)$, $I_{HL}(t)$, and $I_{LH}(t)$. In the HH example, this is given by

$$CCC_{HH,I} = \frac{\langle I_w(t) I_{HH}(t) \rangle}{I_w I_{HH}}, \tag{8}$$

where $CCC_{HH,I}$ is the cross-correlation coefficient between Eve's simulated wire current $I_{HH}(t)$ in the HH case and the *measured* channel current $I_w(t)$.

Finally, similarly to the above voltage correlation attack, Eve guesses that the bit situation is the one with the highest cross-correlation between the current in the wire and in Eve's probing system.

For the channel power protocol, Eve has the measured $U_w(t)$ and $I_w(t)$ at her disposal. She uses her noises to simulate the four possible waveforms, $P_{HH}(t)$, $P_{LL}(t)$, $P_{HL}(t)$, and $P_{LH}(t)$, for the instantaneous power flow from Alice to Bob, and cross-correlates them with the actual measured $P_w(t)$. In the HH example:

$$CCC_{HH,P} = \frac{\langle P_w(t) P_{HH}(t) \rangle}{P_w P_{HH}}. \tag{9}$$

where $P_{HH}(t)$ is the simulated channel power in the HH case, and the measured power is

$$P_w(t) = U_w(t) I_w(t). \tag{10}$$

Finally, Eve guesses that the bit situation is the one with the highest cross-correlation between the power flows in the wire and in her probing system.

### 2.1.2. *Cross-correlation attack directly utilizing Alice's/Bob's and Eve's voltage sources*

There is an alternative way for correlation attacks. Eve measures the wire voltage $U_w(t)$ and wire current $I_w(t)$ [2, 3]. Then, from $I_w(t)$, she uses Ohm's law to calculate the hypothetical voltage drops on Alice's/Bob's possible choices of resistances $R_H$ and $R_L$. From that voltage drop, by using Kirchhoff's loop law, Eve calculates Alice's/Bob's hypothetical noise voltage amplitudes. With these data, she tests four hypotheses:

Hypothesis (i):    Alice has chosen $R_\mathrm{L}$
Hypothesis (ii):   Alice has chosen $R_\mathrm{H}$.
Hypothesis (iii):  Bob has chosen $R_\mathrm{L}$
Hypothesis (iv):   Bob has chosen $R_\mathrm{H}$.

With Hypothesis (i), to utilize Kirchhoff's loop law, using the same current direction (Alice => Bob) as with the power calculation above (Equation 9), Eve takes the sum

$$U_\mathrm{L,A}^*(t) = U_\mathrm{w}(t) + I_\mathrm{w}(t)R_\mathrm{L} \tag{11}$$

to calculate the hypothetical value $U_\mathrm{L,A}^*(t)$ of Alice's noise $U_\mathrm{L,A}(t)$.

To test Hypothesis (i) she finds the cross-correlation between $U_\mathrm{L,A}(t)$ (see Figure 2) and the $U_\mathrm{L,A}^*(t)$ determined by (Equation 11), given by

$$CCC_\mathrm{L,A} = \frac{\left\langle U_\mathrm{L,A}^*(t)U_\mathrm{L,A}(t)\right\rangle}{U_\mathrm{L,A}^* U_\mathrm{L,A}}. \tag{12}$$

Then, with the same equation, she finds the cross-correlation coefficient between $U_\mathrm{L,A}^*(t)$ and $U_\mathrm{H,A}(t)$. If the former result yields the higher cross-correlation, then Eve has determined that Hypothesis (i) is correct; otherwise, Hypothesis (ii) is valid.

Similarly, at Bob's side, with Hypothesis (iii), Eve takes the difference

$$U_\mathrm{L,B}^*(t) = U_\mathrm{w}(t) - I_\mathrm{w}(t)R_\mathrm{L} \tag{13}$$

to calculate the hypothetical value $U_\mathrm{L,B}^*(t)$ of Bob's noise $U_\mathrm{L,B}(t)$.

To test Hypothesis (iii) she takes the cross-correlation coefficient between $U_\mathrm{L,B}(t)$ and the $U_\mathrm{L,B}^*(t)$ determined by (Equation 13), given by

$$CCC_{LB} = \frac{\left\langle U_{LB}^{*}(t)U_{LB}(t)\right\rangle}{U_{LB}^{*}U_{LB}}. \tag{14}$$

Then, she also finds the cross-correlation between $U_{LB}^{*}(t)$ and $U_{H,B}(t)$. If the former result yields the higher cross-correlation, then Eve has determined that Hypothesis (iii) is correct; otherwise, Hypothesis (iv) is valid.

### 2.2. *Unilateral knowledge[3]*

Eve has partial knowledge of only Alice's noises, that is, the noise generator outputs of Alice's resistors, $U_{L,A}(t)$ and $U_{H,A}(t)$. Bob's generator voltages are completely unknown to her. The attack methods described in section 2.1 work even here with a minor modification as described below.

#### 2.2.1. *Cross-correlation attack utilizing Alice's/Bob's and Eve's channel voltages, currents and power*

Eve generates two independent "dummy" thermal noises to substitute for Bob's unknown noise voltages $U_{H,B}(t)$, and $U_{L,B}(t)$. Then she uses the same protocol as the bilateral case (see Section 2.1): she measures $U_w(t)$ and $I_w(t)$, determines $P_w(t)$ from (Equation 9), and uses (Equations 7, 8, and 10) to find the cross-correlation between her four simulated $U_w(t)$, $I_w(t)$, and $P_w(t)$, and the measured $U_w(t)$, $I_w(t)$, and $P_w(t)$.

#### 2.2.2. *Cross-correlation attack utilizing Alice's and Eve's voltage sources*

With $U_{L,A}(t)$ and $U_{H,A}(t)$ partially known, Eve uses a protocol corresponding to the bilateral case (see Section 2.1): she measures $U_w(t)$ and $I_w(t)$ [2, 3]. Then, from $I_w(t)$, she calculates the hypothetical voltage drops on Alice's possible choice of resistances $R_H$ and $R_L$. With these data, she tests two hypotheses:

Hypothesis (i):  Alice has chosen $R_L$.
Hypothesis (ii):  Alice has chosen $R_H$.

Note that, in the bilateral case, she had knowledge of Bob's noises, thus she could form four hypotheses. While in the unilateral case, she has knowledge only Alice's generator voltages, therefore she can form only two hypotheses.

With Hypothesis (i), Eve uses (Equation 11) to find $U_{L,A}^{*}(t)$. Then, she uses (Equation 12) to find the cross-correlation between $U_{L,A}(t)$ and $U_{L,A}^{*}(t)$, and then the cross-correlation between $U_{L,A}^{*}(t)$ and $U_{H,A}(t)$. If the former result yields the higher cross-correlation, then Eve has determined that Hypothesis (i) is correct; otherwise, Hypothesis (ii) is valid.

The extra step that is needed to add to the protocol described in 2.1.2 is as follows: Eve evaluates the measured mean-square voltage on the wire over the whole bit exchange period. From that value, by using (Equation 5), she evaluates the parallel resultant $R_P$ of the resistances of Alice and Bob. From $R_P$ and $R_A$, she calculates $R_B$ by (Equation 6), thus she has cracked the KLJN scheme [3].

## 3. Demonstration

### 3.1. *Noise generation*

#### 3.1.1. *Johnson noise emulation[3]*

First, we generated Gaussian band-limited white noise (GBLWN). Precautions were used to avoid aliasing errors, improve Gaussianity, and reduce bias:

(i)      At first, using the MATLAB randn() function, 224 or 16,777,216 Gaussian random numbers were generated.

(ii)      This process was repeated 10 times to generate 10 independent raw noise series, and then an ensemble average was taken out of those 10 series to create one single noise time function with improved Gaussianity and decreased short-term bias.

(iii)      Then this time series was converted to the frequency domain by a Fast Fourier Transformation (FFT). To get rid of any aliasing error, we opened the real and imaginary portions of the FFT spectrum and doubled their frequency bandwidths by zero padding to represent Nyquist sampling.

(iv)      Finally, we performed an inverse FFT (IFFT) of the zero-padded spectrum to get the time function of the anti-aliased noise. The real portion of the IFFT result is the band-limited, antialiased noise with improved Gaussianity and decreased bias.

The probability plot of the generated noise is shown in Fig. 4, showing that the noise is Gaussian. Fig. 5 demonstrates that the noise has a band-limited, white power density spectrum and that it is anti-aliased.

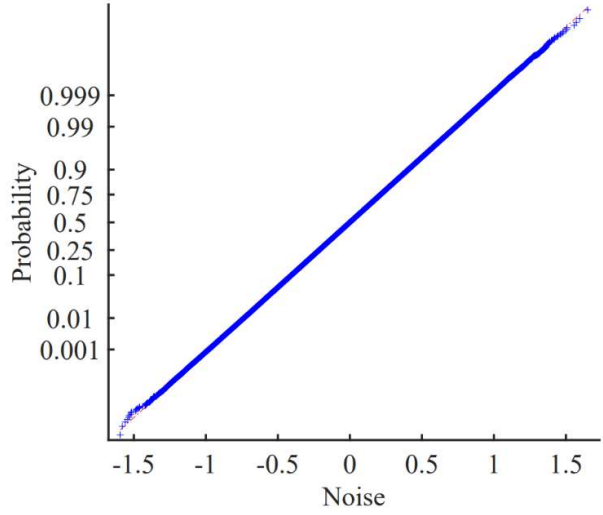

Fig. 4. Normal-probability plot of the noise [3]. A straight line indicates a pure Gaussian distribution.

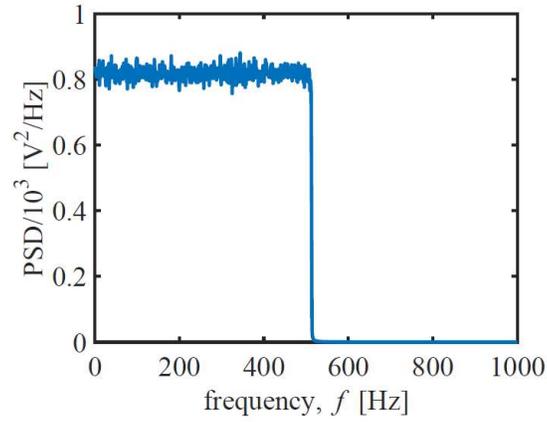

Fig. 5. Power spectral density of the noise [3]. The bandwidth of the noise is 500 Hz, see (5).

From the Nyquist Sampling Theorem,

$$t = \frac{1}{2\Delta f_{\text{B}}},$$ (11)

where $\tau$ represents the time step, an $\Delta f_{\text{B}}$ of 500 Hz renders a time step of $10^{-3}$ seconds.

The final step was to scale this normalized Gaussian noise to the required level of Johnson noise for the given resistance, bandwidth and temperature values ($R_L$, $R_H$, $\Delta f_B$ and $T_{eff}$, respectively). We chose $R_L = 10$ k$\Omega$, $R_H = 100$ k$\Omega$, $\Delta f_B = 500$ Hz and $T_{eff} = 10^{18}$ K).

### 3.1.2. *Setup of Eve's noises*

To generate Eve's noises that are correlated with the noises in the KLJN system, independent "thermal" noises (see Section 3.1.1) are added to the noises of Alice and Bob, and the resulting noises are normalized to have the same effective value as the original noises.

For example, the new correlated noise of Eve corresponding to Alice's thermal noise $U_{L,A}(t)$ is

$$U_{EL,A}(t) = U_{L,A}(t) + M U^*_{EL,A}(t), \tag{15}$$

where $U_{EL,A}(t)$ is the new correlated noise, $U^*_{EL,A}(t)$ is an independently generated "thermal" noise for resistance value $R_L$, and the $M$ coefficient controls the cross-correlation between Eve's and Alice's noise. Note, $U_{EL,A}(t)$ and $U^*_{EL,A}(t)$ have the same effective value. Accordingly,

$$U^*_{EL,A}(t) = x(t)\sqrt{4kT_{eff}R_L\Delta f_B}, \tag{16}$$

where $x(t)$ represents a new noise with 1 Volt effective value.

However, the effective value of $U_{EL,A}(t)$ does not satisfy the Johnson formula anymore, so in order to call it a thermal noise, it must be scaled to the proper level [3]:

$$U_{L,A}(t) = \frac{U^*_{EL,A}(t)}{\sqrt{\left\langle\left[U^*_{EL,A}(t)\right]^2\right\rangle}}\sqrt{4kT_{eff}R_L\Delta f_B}, \tag{17}$$

In this way, the same $M$ used for the different resistance choices of Alice and Bob results in the *same* cross-correlation coefficient between Eve's noises and the corresponding noises of the communicating parties. Thus, four new sets of independent additive noises have been generated using the same method as in Section 3.1.

A realization of Eve's noise voltages in comparison to Alice's and Bob's noise voltages over 100 milliseconds is displayed in Fig. 6. $U_{H,A}(t)$ is the noise voltage of Alice's $R_H$, $U_{L,A}(t)$ is the noise voltage of Alice's $R_L$, $U_{H,B}(t)$ is the noise voltage of Bob's $R_H$, and $U_{L,B}(t)$ is the noise voltage of Bob's $R_L$ (see Fig. 2). Each time step is one millisecond.

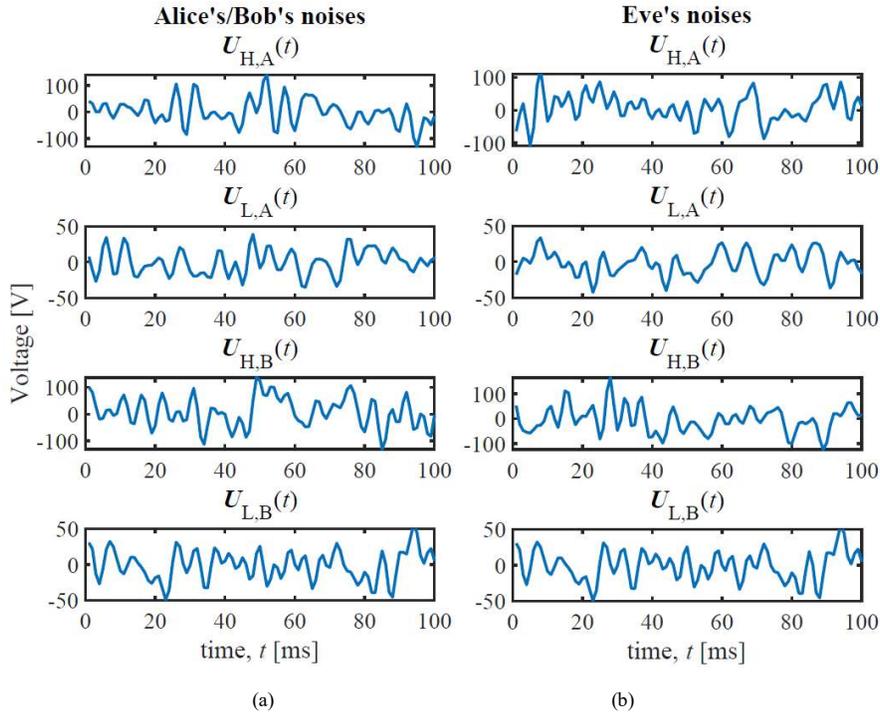

Fig. 6. A realization of $U_{H,A}(t)$, $U_{L,A}(t)$, $U_{H,B}(t)$, and $U_{L,B}(t)$ (see Fig. 2) for Alice and Bob (a), and for Eve (b), displayed over 100 milliseconds.

### 3.2.  *Attack demonstration when Eve knows both noises (bilateral attacks)*

#### 3.2.1.  *Bilateral attack demonstration utilizing cross-correlations between Alice's/Bob's and Eve's wire voltages, currents and powers*

Computer simulations were executed with MATLAB, and each simulation was run 1000 times, see Tables 1-4. During the attacks, the sample size (number of time steps) is 1000, and the KLJN system is kept in the LH state. A realization of the wire voltage, current, and power, $U_w(t)$, $I_w(t)$, and $P_w(t)$, under the LH condition over 100 milliseconds is displayed in Fig. 7.

The results of the attack are shown in Table 1. The HL situation of Eve has zero cross-correlation with Eve's probe signals because, in this case, Alice/Bob, and Eve do not have a common noise. In all the other cases, they have at least one mutual noise. In the LH attack situation both noises are common thus they yield the highest cross-correlation with Eve's LH probe, therefore Eve guesses that the secure resistance situation is LH.

Using the correlations between the voltages for the attack yielded the highest probability $p$ of successful guessing, and the correlations between the powers yielded the lowest one. This is because power is the product of voltage and current (see Equation 9), and the inaccuracies in both the voltage and current correlations affect the power correlations.

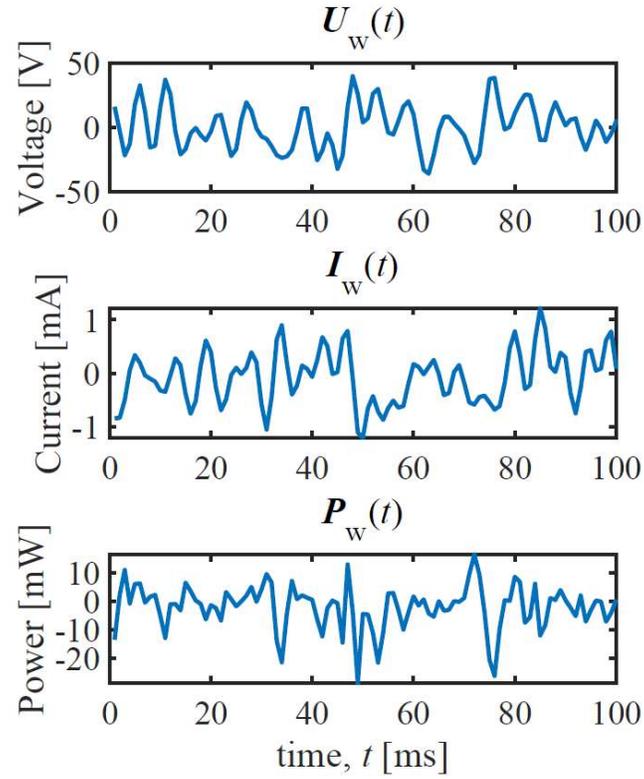

Fig. 7. A realization of $U_w(t)$, $I_w(t)$, (see Fig. 2) and $P_w(t)$ (see Equation 9) for the LH situation is displayed over 100 milliseconds [3]. Eve measures and records these data.

Table 1. Simulation of the average cross-correlation coefficient, *CCC* (see Equations 7, 8, and 10), and Eve's $p$ of correctly guessing the LH bit situations at varying multipliers $M$ (see Section 3.1.2). As $M$ increases, which implies increasing differences between the noises of Alice/Bob and Eve's probing noises, the cross-correlation decreases. Yet, in each case of the present situation, the LH case yields the highest correlation. Thus, Eve guesses that LH is the secure bit situation. The correlations $CCC_u$ between the voltages yielded $p_u$, which had the highest probability $p$ of successful guessing, and the correlations $CCC_p$ between the powers yielded the lowest $p$. This is because power is the product of voltage and current (see Equation 9), and the inaccuracies in both the voltage and current correlations affect the power correlations.

| Eve's probing bit | $M$ | $CCC_u$ | $p_u$ | $CCC_i$ | $p_i$ | $CCC_p$ | $p_p$ |
|---|---|---|---|---|---|---|---|
| HH | 0 | 0.213960 | | 0.674280 | | 0.286570 | |
| LL | | 0.675060 | | 0.212460 | | 0.285320 | |
| HL | | 0.002088 | | 0.000370 | | 0.000157 | |
| LH | | 1 | 1 | 1 | 1 | 1 | 1 |
| HH | 0.1 | 0.039932 | | 0.109800 | | 0.009254 | |
| LL | | 0.347620 | | 0.125890 | | 0.076979 | |
| HL | | 0.000239 | | -0.000177 | | -0.000988 | |
| LH | | 0.485410 | 1 | 0.216720 | 1 | 0.114570 | 0.995 |
| HH | 0.5 | 0.009068 | | 0.026372 | | -0.001256 | |
| LL | | 0.080356 | | 0.026207 | | 0.003574 | |
| HL | | -0.000313 | | 0.001804 | | -0.000401 | |
| LH | | 0.112630 | 0.998 | 0.044934 | 0.781 | 0.003888 | 0.550 |
| HH | 1 | 0.004431 | | 0.012502 | | -0.000065 | |
| LL | | 0.040189 | | 0.012589 | | 0.000728 | |
| HL | | -0.000166 | | -0.000217 | | -0.000023 | |
| LH | | 0.056232 | 0.904 | 0.022431 | 0.651 | 0.001271 | 0.544 |
| HH | 1.5 | 0.001820 | | 0.008615 | | -0.000281 | |
| LL | | 0.027865 | | 0.007432 | | -0.001454 | |
| HL | | -0.000490 | | 0.000293 | | -0.000130 | |
| LH | | 0.038744 | 0.827 | 0.014521 | 0.6 | 0.001894 | 0.524 |
| HH | 10 | 0.000908 | | 0.000263 | | 0.000477 | |
| LL | | 0.003666 | | 0.001263 | | 0.001151 | |
| HL | | -0.000805 | | -0.000011 | | -0.000055 | |
| LH | | 0.006126 | 0.558 | 0.002058 | 0.528 | 0.001359 | 0.518 |

### 3.2.2. *Bilateral attack demonstration utilizing cross-correlations among the voltage sources*

Fig. 8 shows the hypothetical noise voltages generated by Eve's simulations across $R_L$ and $R_H$ (see Equation 11) [3].

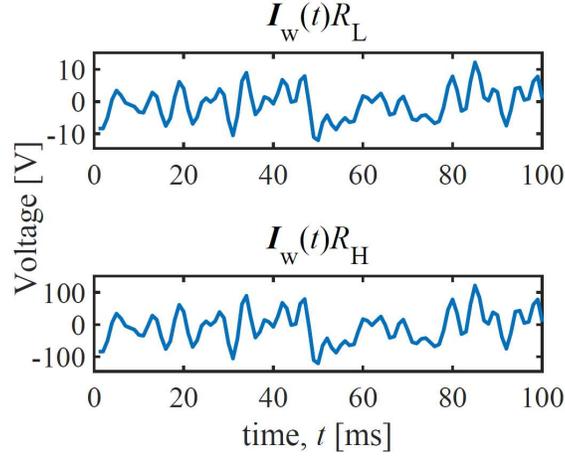

Fig. 8. Hypothetical noise voltage drops across $R_L$ and $R_H$ by Eve's current measurements and Ohm's law. These results are used in (Equations 11 and 13) to calculate the hypothetical waveforms for $U_{L,A}(t)$, $U_{H,A}(t)$, $U_{L,B}(t)$, and $U_{H,B}(t)$.

With this information, Eve uses (Equation 11) to find $U_{L,A}^*(t)$, (Equation 12) to find the cross-correlation between $U_{L,A}^*(t)$ and $U_{L,A}(t)$, (Equation 13) to find $U_{L,B}^*(t)$, and (Equation 14) to find the cross-correlation between $U_{L,B}^*(t)$ and $U_{L,B}(t)$. A simulation results of the cross correlation for each bit case is displayed in Table 2. As the multiplier $M$ in the superposition increases (see Section 3.1.2), the correlations decrease. Yet in each case at the present (LH) situation, the highest correlation ended up being the LH case, thus Eve has decided that LH is the secure resistance situation.

Table 2. Simulation of the cross-correlation coefficients *CCC* at the attack described in Section 2.1.2. (see Equations 12 and 14), and Eve's correct-guessing probability, *p*, at varying multipliers *M* (see Section 3.1.2). As *M* increases, which implies increasing differences between the noises of Alice/Bob and Eve's probing noises, the cross-correlation decreases. Yet, in each case of the present situation, the $R_L$ hypothesis yields the highest cross-correlation with Alice's noise. Thus, Eve guesses that Alice has $R_L$. The same procedure done for Bob's noise results in the highest cross-correlation for the $R_H$ hypothesis at Bob's side.

| $M$ | $CCC_{L,A}$ ($R_L$) | $CCC_{L,A}$ ($R_H$) | $CCC_{L,B}$ ($R_L$) | $CCC_{L,B}$ ($R_H$) | $p$ |
|---|---|---|---|---|---|
| 0 | 1 | -0.015600 | 0.00028951 | 0.560400 | 1 |
| 0.1 | 0.515040 | 0.008990 | 0.002046 | 0.131430 | 0.992 |
| 0.5 | 0.118910 | -0.028922 | 0.000017 | 0.029605 | 0.669 |
| 1 | 0.060229 | 0.009445 | -0.000460 | -0.011328 | 0.569 |
| 1.5 | 0.048146 | 0.039420 | 0.000012 | 0.026396 | 0.547 |
| 10 | 0.007059 | -0. 001521 | 0.000348 | -0.011921 | 0.501 |

### 3.3. *Attack demonstration when Eve knows only one of the sources (unilateral attacks)*

#### 3.3.1. *Unilateral attack demonstration utilizing cross-correlations between Alice's/Bob's and Eve's wire voltages, currents and powers*

Like the bilateral results (see Section 3.2.1), the HL situation of Eve has zero cross-correlation with Eve's probe signals because, in this case, Alice/Bob, and Eve do not have a common noise. In all the other cases, they have at least one mutual noise. In the LH probe attack both noises are common thus they yield the highest cross-correlation with Eve's LH probe, therefore Eve guesses that the secure resistance situation is LH, see Table 3.

Using the correlations between the voltages for the attack yielded the highest probability $p$ of successful guessing, and the correlations between the powers yielded the lowest one. This is because power is the product of voltage and current (see Equation 9), and the inaccuracies in both the voltage and current correlations affect the power correlations.

Table 3. Simulation of the cross-correlation coefficient, *CCC* (see Equation 9), and Eve's $p$ of correctly guessing the LH bit situations at varying multipliers $M$ (see Section 3.1.2). As $M$ increases, which implies increasing differences between the noises of Alice/Bob and Eve's probing noises, the cross-correlation decreases. Yet, in each case of the present situation, the LH case yields the highest correlation. Thus, Eve guesses that LH is the secure bit situation. The correlations $CCC_u$ between the voltages yielded $p_u$, which had the highest probability $p$ of successful guessing, and the correlations $CCC_p$ between the powers yielded the lowest $p$. This is because power is the product of voltage and current (see Equation 9), and the inaccuracies in both the voltage and current correlations affect the power correlations.

| Eve's probing bit | $M$ | $CCC_u$ | $p_u$ | $CCC_i$ | $p_i$ | $CCC_p$ | $p_p$ |
|---|---|---|---|---|---|---|---|
| HH | | -0.000023 | | 0.000145 | | 0.000005 | |
| LL | 0 | 0.673820 | | 0.090397 | | 0.16462 | |
| HL | | -0.001374 | | -0.000015 | | -0.000032 | |
| LH | | 0.909330 | 1 | 0.211650 | 0.98 | 0.285270 | 0.992 |
| HH | | -0.001021 | | 0.000957 | | -0.000044 | |
| LL | 0.1 | 0.379860 | | 0.048214 | | 0.043738 | |
| HL | | 0.000501 | | 0.000236 | | 0.001270 | |
| LH | | 0.468500 | 1 | 0.110610 | 0.863 | 0.076680 | 0.801 |
| HH | | -0.000207 | | 0.000175 | | -0.0013809 | |
| LL | 0.5 | 0.079899 | | 0.001031 | | 0.000677 | |
| HL | | -0.000998 | | 0.009610 | | -0.0011319 | |
| LH | | 0.108640 | 0.994 | 0.024430 | 0.581 | 0.002906 | 0.518 |
| HH | | 0.000288 | | 0.000091 | | 0.001396 | |
| LL | 1 | 0.040892 | | 0.012783 | | -0.000151 | |
| HL | | 0.000892 | | -0.000116 | | 0.001294 | |
| LH | | 0.054347 | 0.886 | 0.057347 | 0.542 | 0.003950 | 0.517 |
| HH | | -0.000615 | | -0.000169 | | -0.000136 | |
| LL | 1.5 | 0.027794 | | 0.008971 | | 0.000731 | |
| HL | | -0.000503 | | 0.000944 | | 0.000212 | |
| LH | | 0.037697 | 0.805 | 0.026552 | 0.523 | 0.001249 | 0.508 |
| HH | | -0.000042 | | -0.000376 | | -0.001033 | |
| LL | 10 | 0.004576 | | 0.000405 | | 0.002124 | |
| HL | | 0.000462 | | -0.000483 | | 0.000807 | |
| LH | | 0.005711 | 0.539 | 0.000949 | 0.514 | 0.002155 | 0.501 |

Finally, Eve evaluates the measured mean-square voltage on the wire over the bit exchange period. From that value, by using (Equation 5), she evaluates the parallel resultant $R_P$ of the resistances of Alice and Bob. From $R_P$ and $R_A$, she can calculate $R_B$. In this particular case, from the mean-square voltage Eve will learn that the actual situation is LH thus Bob has chosen $R_H$ because Alice has $R_L$ [3].

### 3.3.2. *Unilateral attack demonstration utilizing cross-correlations among the voltage sources*

Eve uses (Equation 11) to find $U_{L,A}^*(t)$ and (Equation 12) to find the cross-correlation between $U_{L,A}^*(t)$ and $U_{L,A}(t)$. A simulation results of the cross correlation for each bit case is displayed in Table 4. As the multiplier $M$ increases (see Section 3.3.1), the correlation decreases, yet the highest correlation always ended up being the LH case, thus Eve has decided that LH is the secure resistance situation.

Table 4. A realization of the average cross-correlation coefficient, *CCC* (see Equation 9), and Eve's correct-guessing probability, *p*, at varying multipliers *M* (see Section 3.3.1). As *M* increases, which implies increasing differences between the noises of Alice/Bob and Eve's probing noises, the cross-correlation decreases. Yet the $R_L$ case yields the highest correlation. Thus, Eve guesses that Alice has $R_L$.

| $M$ | $CCC_{L,A}(R_L)$ | $CCC_{L,A}(R_H)$ | $p$ |
|-----|-----|-----|-----|
| 0 | 1 | -0.015600 | 1 |
| 0.1 | 0.515220 | 0.008990 | 1 |
| 0.5 | 0.118910 | -0.028922 | 0.999 |
| 1 | 0.060047 | 0.009445 | 0.907 |
| 1.5 | 0.048146 | 0.040953 | 0.804 |
| 10 | 0.005654 | -0.001521 | 0.546 |

Finally, Eve evaluates the measured mean-square voltage on the wire over the bit exchange period. From that value, by using Equation 5, she evaluates the parallel resultant $R_P$ of the resistances of Alice and Bob. From $R_P$ and $R_A$, she can calculate $R_B$. In this particular case, from the mean-square voltage Eve will learn that the actual situation is LH thus Bob has chosen $R_H$ because Alice has $R_L$ [3].

## 4. Conclusion

Secure key exchange protocols utilize random numbers, and compromised random numbers lead to information leak. So far, it had been unknown how Eve can utilize partial knowledge about the random number generators to attack the KLJN protocol. We explored various situations how RNGs compromised by statistical knowledge can be utilized by Eve.

We have introduced four new attacks against the KLJN scheme. The defense against these attacks is the usage of true random number generators (with proper tamper resistance) at Alice's and Bob's sides.

Finally, it is important to note that this exploration was done assuming an ideal KLJN scheme. Future work would involve a practical circuit implementation or a cable simulator and related delays and transients.